\documentclass[twocolumn,aps,prb,floatfix]{revtex4-2}

\usepackage{amsmath}
\usepackage{amsfonts}
\usepackage{graphicx}
\usepackage{appendix}
\usepackage{dsfont}
\usepackage{amssymb}
\usepackage{bm}
\usepackage{xcolor}
\usepackage{ulem}
\usepackage{soul}
\usepackage{mathrsfs}
\usepackage{amsmath}
\usepackage{esint}
\usepackage{natbib}
\usepackage{braket}
\usepackage{nccmath}
\usepackage{array}
\newcommand{\beq}{\begin{equation}}
\newcommand{\eneq}{\end{equation}}
\newcommand{\be}{\begin{equation}}
\newcommand{\ee}{\end{equation}}
\newcommand{\bea}{\begin{eqnarray}}
\newcommand{\eea}{\end{eqnarray}}

\usepackage[final]{hyperref} 
\hypersetup{
	colorlinks=true,       
	linkcolor=blue,        
	citecolor=blue,        
	filecolor=magenta,     
	urlcolor=blue         
}

\usepackage{multirow}
\usepackage{wasysym}

\begin{document}

\title{Charge current and phase diagram of the disordered open longer-range
Kitaev chain}
 
\author{Emmanuele G. Cinnirella$^{(1)}$, Andrea Nava$^{(2)}$, and   Domenico Giuliano$^{(1,2)}$}
\affiliation{
$^{(1)}$ Dipartimento di Fisica, Universit\`a della Calabria Arcavacata di 
Rende I-87036, Cosenza, Italy and \\ I.N.F.N., Gruppo collegato di Cosenza 
Arcavacata di Rende I-87036, Cosenza, Italy \\
$^{(2)}$ Institut f\"ur Theoretische Physik, Heinrich-Heine-Universit\"at, 40225 D\"usseldorf, Germany  }
 
\begin{abstract}

We compute the disorder averaged dc conductance in the non-equilibrium steady 
state that sets in between a longer-range Kitaev chain and
a metallic lead connected to an external reservoir, as a function of the system parameters
and of the disorder strength. From our results, we map out the phase diagram of the 
disordered  chain for different types of disorder  and discuss the corresponding effects of the 
interplay between topology and disorder in the system. To do so, 
we set up a  combined analytical and numerical approach, 
which is potentially amenable of straightforward generalizations to other disordered topological systems.  

\end{abstract}
\date{\today}
\maketitle

\section{Introduction}
\label{intro}

Recently, one-dimensional (1D) electronic  topological systems have become the
subject of an intensive and systematic study, primarily due to 
their remarkable property of developing (topological) phases characterized by 
emerging, low-lying, charge-neutral modes, localized at the endpoint(s) of 
the system \cite{Hasan2010,Qi2011}. Indeed, the peculiar properties of the localized
modes, in particular their robustness against fluctuations in the system operating parameters
(including the temperature), disorder, and so on, make them particularly suitable to 
engineer robust qubits and fault tolerant quantum computing protocols. The emergence of topological phases 
can be realized in a wide class of solid-state systems \cite{Moore1991,Nayak2008,Oreg2010,Mousavi2015,Chen2016}, 
and has also recently been 
experimentally seen in optical simulations of topological systems \cite{Cardano2017,Derrico2020}. 
  
An effective way to probe  topological phases for given system parameters is 
to compute a topological invariant quantity, such as the Zak phase (the ``winding number'')
 $\omega$ \cite{Zak1989}.
 $\omega$ is computed in models with periodic boundary conditions: it takes either 0, or nonzero, integer values,
 and its specific value counts how many independent localized modes emerge at each endpoint of 
 the system with open boundary conditions \cite{Fidkowski2011,Chen2020}.  Therefore, when $\omega = 0$ the system
 lies within a topologically trivial phases, while topological non-trivial phases are signaled by a nonzero value of 
 the topological invariant. Electronic models with short range electron hopping and/or interactions, 
 typically host at most one single localized mode at endpoints of the system, which implies $| \omega | = 0,1$,
 while  models with longer (that is, beyond nearest neighbor) hopping and/or interactions may develop phases
 with $|\omega |>1$. Such ``higher-$\omega$'' 
 phases are expected to emerge in solid-state systems, both normal \cite{Li2014,Li2019,Malakar2023,Perez2019}, and superconducting
 \cite{Niu2012,Degottardi2013,Lieu2018,Li2021}, as well as in optical systems  \cite{Cardano2017,Derrico2020}.  
 Thus, it is important to investigate their properties in 
 pertinently constructed ``minimal'' prototypical models.
 
 To formally describe realistic topological systems one has to account for the unavoidable effect of disorder.
In this respect, the interplay of disorder and topology has several, nontrivial consequences on 
the phase diagram of a disordered topological model, with features depending on the model 
Hamiltonian of the topological system in the ``clean'' limit, as well as on the specific 
kind of disorder one considers, together with the related symmetries underneath the disorder
potential. For instance,  in topological systems with only short range 
electron hopping strengths and interactions (including pairing interaction in 
superconducting models), a small amount of Anderson-like 
disorder may work to enforce the stability of the topological phase, 
which may ``reenter'' into regions of values of the Hamiltonian parameters that
would be topologically trivial  in the clean limit 
\cite{Zuo2022,Li2009,Cinnirella2024,Pientka2013,Nava2017,Cinnirella2025}.
 Also,  the effects of 
 adding disorder to the model Hamiltonian strongly depend on 
  whether the disorder preserves, or not, some specific symmetries of the 
 system Hamiltonian \cite{Liu2022,Nava2023}. Given their richer phase diagram, compared to short range
 models, it is therefore quite interesting to address the effects of the interplay between
 disorder and topological phases in systems with more than one topological phase in
 their phase diagram.
 
 In this paper, we study the phase diagram of the longer range 1D Kitaev Hamiltonian (2LRK):
 a model of spinless electrons 
 with both first- and second-neighbor single electron hopping terms and first- and 
 second-neighbor pairing terms \cite{Niu2012}, in the presence of three different kinds of disorder potentials. 
 Specifically, we consider the case of  Anderson disorder, of  the Aubry-Andr\'e-Harper model 
\cite{Harper1955,Aubry1980}, and of a disorder affecting the hopping and the pairing strengths in the model
Hamiltonian. In all three cases we have to define  a pertinent generalization of 
$\omega$, in order to properly reconstruct the corresponding topological
phases in the phase diagram. To do so, we employ an adapted version of the approach used 
in \cite{Nava2023,Cinnirella2024,Cinnirella2025}. Specifically, we focus onto an open 2LRK, connected by one
of its sites to a metallic lead. On its own, the lead is connected to an 
external thermal bath, whose interplay with the system dynamics drives its time evolution toward a pertinent 
non-equilibrium steady state (NESS), in which a potential bias between the lead and the 2LRK induces a current flow
across the interface between the two of them. Estimating, for each single realization of the disorder, 
 the corresponding dc conductance, $G_{{\rm int},n}$, for each site of the 
chain and summing over all the sites, we obtain a quantity given by a prefactor depending only on 
the total Hamiltonian parameters in the clean limit, times $\omega$ (which is always quantized), 
computed for that realization of the disorder. 
Eventually, by averaging over several realizations of disorder, we construct the corresponding
 extension of $\omega$, from which we infer the phase diagram of the disordered system. 

Following the method of Refs.\cite{Nava2023,Cinnirella2024,Cinnirella2025}, we
 describe the time evolution of the open system toward the NESS by means of  Linblad Master
equation (LME) approach \cite{Lindblad1976}, which has already been successfully applied  
 to study the onset of  dynamical  topological phase transitions  
 \cite{Diehl2011,Goldstein2019,Shavit2020,Cui2019,Nava2023_s1,Nava2023_s2}, as well as
 the NESS that sets in the system as a consequence of the coupling to the external reservoir
 \cite{Deleeuw2024,Tarantelli2021,Pizorn2013,Benenti2009,Benenti2009_2,Nava2021L}.   
 Within LME approach, we recover the time evolution equations for the two-fermion correlation
 matrix elements, in terms of which we express the dc conductance at each site. In the
 $t \to \infty$ limit, we obtain the correlation matrix elements and, therefore, the dc conductances
 $G_{{\rm int},n}$ in the NESS. Remarkably, while analytically solving the LME to recover the correlation matrix
 at any $t$ is practically impossible, by pertinently adapting the formalism developed in Ref.\cite{Guimaraes2016}, we are 
 able to provide an analytical expression  for $G_{\rm int}$
 in the weak coupling limit in both the coupling between the 
 chain and the lead and between the lead and the reservoir, at any given realization of the disorder. 
 Eventually, we numerically average our results for $G_{\rm int}$ (and, therefore, for $\omega$) 
 over all the realizations of the disorder. Plotting the corresponding average
as a function of the tuning Hamiltonian parameter(s) and of the disorder 
strength $W$, we reconstruct the whole phase diagram of the disordered system. 

At a synoptic look at our results, we evidence how different kinds of disorder affect
the topological phases in the 2LRK. In particular, while we confirm that the uncorrelated 
(Anderson) disorder enforces (in the weak disorder limit) the $|\omega|=1$ topological phase,
we also note that it strongly suppresses phases with $|\omega| = 2$. At variance, 
a correlated disorder, can tend to enforce higher-$|\omega|$ topological 
phases, eventually generalizing to those phases the reentrant behavior of  the 
$|\omega|=1$ phase in the presence of Anderson disorder. 
Throughout our derivation, we evidence how, given its generality and the complementary use of analytical 
and numerical methods, our approach has a potential range of applicability much wider than the specific cases we
analyze here. Moreover,  in view of its potential potential experimental implementation 
\cite{Maiellaro2019},  it can likely provide a systematic tool to investigate the intriguing interplay between 
topology and disorder in solid state electronic systems. 
 
The paper is organized as follows:

\begin{itemize}

\item In Section \ref{modham} we present  the model Hamiltonian for the  isolated 2LRK, together with  the various types of 
disorder that we consider.  Then, we provide the formal description of the superconducting chain 
connected to the lead which, on its own, is connected to the external reservoir. Eventually, we review the LME 
approach and its application to the specific system that we consider here.
 
 \item In Section \ref{cness} we present and discuss our results for the phase diagram of the open 2LRK derived from 
 the dc conductance  in the NESS, $G_{\rm int}$.
 
 \item In Section \ref{concl} we summarize our results and provide some possible further developments of our work.
 
\item In Appendix \ref{current_o} we provide  the details on the mathematical derivation of the dc conductance 
through the interfaces between the Kitaev chain and the metallic  leads.

\end{itemize}
 
\section{System Model Hamiltonian}
\label{modham}

In this Section we provide the model Hamiltonian for the 2LRK, first in the clean limit, and 
then in the presence of  disorder.   Eventually, we add a coupling between the Kitaev chain and
the metallic lead connected to the external reservoir. Due to the coupling to the reservoir, 
the superconducting chain and the lead, taken all together,  constitute an open system, which we formally treat 
within  LME approach. Therefore, we devote the last part of this Section to 
review LME formalism  and how to apply it to our specific system. In Appendix \ref{current_o} we 
present the mathematical derivation of the electric current through the NS interfaces between the 
chain and the lead. 

\subsection{Model Hamiltonian for the longer-range Kitaev chain}
\label{sub1_modham}

Kitaev lattice model for a one-dimensional p-wave superconductor has by now become a
prototypical system hosting Majorana real fermionic modes localized at the boundaries of a finite chain \cite{Kitaev2001},
also in view of the various possible platform that have been proposed as potential realizations of 
the model \cite{Oreg2010,Lutchyn2010,Sau2012,Dvir2023}.  Moreover, the Kitaev model is 
well-known to provide a fermionic description of the one-dimensional quantum Ising model, with the latter model
being mapped onto the former via Jordan-Wigner fermionization procedure  (see, for instance, 
\cite{Chhaied2021} and references therein). 
Specifically, tuning the chemical potential of the system, it is possible to drive it across a topological phase 
transition between a topological gapped phase, in which the localized Majorana modes emerge at the system
boundaries, and a trivial gapped phase, in which no modes are present within the gap. 
Generalizations of the  Kitaev Hamiltonian   characterized by both normal single electron hopping, as well as pairing,  
ranging beyond  nearest-neighboring sites only,  have been discussed in the recent literature 
\cite{Niu2012,Degottardi2013,Vodola2014,Viyuela2016,Lepori2016,Giuliano2018,Li2021}, also in order to 
make the model Hamiltonian for the system better fit what one should expect from realistic systems 
\cite{Pientka_2013,Pientka2014}. In fact, allowing for longer-range fermion hopping/pairing allows for the emergence 
of novel phases in the  phase diagram of the Kitaev chains, such as multi-Majorana phases with several
real fermionic zero-modes at the edges of the chain, or phases in which long-range correlations pair the edge Majorana modes
in massive Dirac boundary modes, et cetera.

\begin{figure}
\includegraphics[width=6cm]{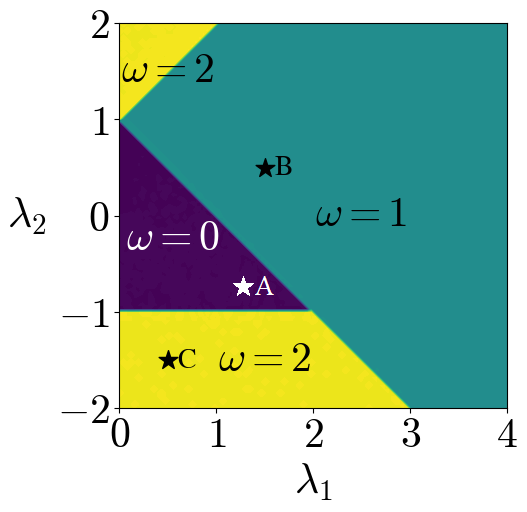}
\caption{Phase diagram of the longer-range Kitaev chain with $g=1$ in the $\lambda_1-\lambda_2$ plane 
\cite{Lieu2018}. The various phases
are evidenced by different colors:  purple for the trivial phase ($\omega =0$), green for the topological phase with 
$\omega=1$, yellow for the topological phase with $\omega=2$. In the following, we focus on the points 
marked with  the $\star$ to provide a first discussion of the effects of adding disorder to the clean system. 
 }
\label{pha_dia}
\end{figure}

In this paper we focus onto the 2LRK, with fermion hopping and pairing taking place between nearest neighboring 
and next-to-nearest neighboring sites. Following Ref.\cite{Niu2012} we  write the model Hamiltonian $H_{2K}$ over
an $L$-site open chain as 

\bea
H_{2K} &=& 2g \sum_{n=1}^L \chi_n^\dagger \chi_n -\lambda_1 \sum_{n=1}^{L-1}
\{\chi_n^\dagger \chi_{n+1}   + \chi_n^\dagger \chi_{n+1}^\dagger +{\rm h.c.} \}  \nonumber \\
& & - \lambda_2 \sum_{n=1}^{L-2} \{\chi_n^\dagger \chi_{n+2} + \chi_n^\dagger \chi_{n+2}^\dagger 
+ {\rm h.c.} \}
\:\: . 
\label{mh.1}
\eea
\noindent
In Eq.(\ref{mh.1}), $\chi_n$ and $\chi_n^\dagger$ are the single-fermion annihilation and creation operators at site-$n$, satisfying the 
anticommutation algebra $\{\chi_n,\chi_{n'}^\dagger \}=\delta_{n,n'}$, $2g$ is the uniform 
chemical potential of the model and, for the sake of simplicity, following \cite{Niu2012}, we have
assumed that the nearest-neighbor single electron hopping and the corresponding pairing strengths are
equal to each other and both equal to $\lambda_1$, while the next-to-neirest single electron hopping and 
pairing strengths are both equal to $\lambda_2$. Finally, h.c. stands for Hermitian conjugate and the 
sums ranges over lattice sites consistently with the range of the corresponding terms and with the open boundary
conditions on the chain. To diagonalize $H_{2K}$,we rewrite it in the Nambu spinor basis as

\beq
H_{2K}   = \sum_{q>0} \: [c_q^\dagger , c_{-q} ] \: [\vec{\sigma}   \cdot \vec{{\cal H}} (q) ] \:\left[\begin{array}{c}
c_q \\ c_{-q}^\dagger \end{array}   \right]
\:\: , 
\label{mh.3}
\eneq
\noindent
with $\vec{\sigma}$ being the triple of the Pauli matrices and 

\beq
\vec{{\cal H}} (q) = 2 \left[ \begin{array}{c} 0 \\ \lambda_1 \sin (q) + \lambda_2 \sin (2q) \\
 g - \lambda_1 \cos (q) - \lambda_2 \cos (2q) 
\end{array}   \right] 
\:\: . 
\label{mh.4}
\eneq
\noindent
The model Hamiltonian in Eq.(\ref{mh.1}) was originally derived in Ref.\cite{Niu2012} by applying 
the Jordan-Wigner fermionization procedure to the three spin extension of the transverse field Quantum
Ising model. Taken as just a fermionic model Hamiltonian for a topological superconductor, 
$H_{2K}$ possesses particle-hole, time-reversal and chiral symmetries: it belongs to 
the BDI class of topological insulators and the topological phases can be characterized in 
terms of an integer-valued topological invariant $\omega$.  The single-particle energy spectrum of $H_{\rm 2K}$ is, 
accordingly,  particle-hole symmetric: at any $q$ there is a pair of associated energy eigenvalues
of opposite sign, $\pm \epsilon_q$, given by 

\beq
 \epsilon_q  =  2 \sqrt{g^2 + \lambda_1^2 + \lambda_2^2 + 2 \lambda_1 (\lambda_2 - g) \cos (q) - 
2 g \lambda_2 \cos (2q) }
\:\: . 
\label{mh.4bis}
\eneq
\noindent
Apparently, Eq.(\ref{mh.4bis}) evidences a fully gapped spectrum, with 
minimum energy gap $\Delta_{2K}$ depending on the system parameters
(for instance, assuming that $g,\lambda_1$ and $\lambda_2$ are all $>0$, for small 
$g$ we obtain $\Delta_{2K}=2 |g-\lambda_1-\lambda_2|$). 
With periodic boundary conditions, the eigenmodes corresponding to the two solutions in Eq.(\ref{mh.4bis}) are
given by 

\beq
\eta_{2K,q,\pm} = \sum_{n=1}^L \: \{[u_{q,n,\pm}]^* \chi_n + [v_{q,n,\pm}]^* \chi_n^\dagger \} 
\;\; , 
\label{mh.4ter}
\eneq
\noindent
with (in Nambu notation)

\bea
\left[\begin{array}{c}   
u_{q,n,+} \\ v_{q,n,+} \end{array}   \right] &=& a_+  \left[\begin{array}{c}   \cos \left(\frac{\xi_q}{2} \right) \\ 
i \sin \left(\frac{\xi_q}{2}   \right) \end{array} \right] e^{iqn}  \;\; , \nonumber \\ \; \left[\begin{array}{c}   
u_{q,n,-} \\ v_{q,n,-} \end{array}   \right] &=& a_- \left[\begin{array}{c} i   \sin \left(\frac{\xi_q}{2} \right) \\ 
\cos \left(\frac{\xi_q}{2}   \right) \end{array} \right] e^{iqn} \;\; , 
\label{mh.4quater}
\eea
\noindent
and $\cos (\xi_q) = (g - \lambda_1 \cos (q) - \lambda_2 \cos (2q))/\epsilon_q$, 
$\sin (\xi_q) = (\lambda_1 \sin (q) + \lambda_2 \sin (2q))/\epsilon_q$, 
 $a_\pm$ being normalization constants. Over the open chain, subgap localized real fermionic modes
may appear at the boundaries of the chain, depending on the specific values of the system parameters.
While, strictly speaking, no boundary modes
emerge in the (closed) chain with periodic boundary conditions, to count the number of  localized real fermionic modes that would be present at
each endpoint of the chain with open boundary condition, in the case of a closed
longer-range Kitaev Hamiltonian and in the absence of disorder one may compute
the Zak phase $\omega$   \cite{Zak1989},   given by

\beq
\omega = \frac{i}{\pi}   \: \int_{\rm BZ} \: d q \: \langle u_q | \partial_q | u_q\rangle 
\:\: , 
\label{mh.5}
\eneq
\noindent
with $\int_{\rm BZ} \: dq $ denoting the integral over the full Brillouin zone (that is, 
$\int_0^{2\pi} \: d q $), and $|u_q\rangle$ being the eigenket corresponding to the 
energy eigenvalue $-\epsilon_q$. In our specific model, we can obtain $\omega = 0$ (that 
corresponds to the topologically trivial phase) or $\omega = 1,2$ (respectively 
corresponding to the topological phase with one and two Majorana mode(s) at each 
edge).  Combining the whole set of values of $\omega$ recovered for different values
of the system parameters, one obtains a phase diagram such as the one we draw in 
Fig.\ref{pha_dia}    by repeating the derivation of Ref.\cite{Lieu2018}, that is, 
by setting $g=1$ and by varying both $\lambda_1$ and $\lambda_2$. 
Just as in \cite{Lieu2018}, we draw the different phases in color,
which allows us to identify at a glance the various phase transition lines, separating 
regions drawn in different colors. Particularly interesting is the possibility to 
pertinently act on the parameters so to realize direct phase transitions between 
phases characterized by any pairs of values of $\omega$ between 0,1 and 2. 
As we evidence in the following, the disorder can strongly modify the topology
of the phase diagram from what we draw in Fig.\ref{pha_dia}. 

Analytically recovering solutions for the Nambu spinors corresponding to 
the Hamiltonian eigenmodes over the chain with open boundary conditions 
requires much more care than in the periodic case (see, for instance, 
Ref.\cite{Giuliano2016} for a similar analysis performed in the case of 
the Jordan-Wigner fermionic representation of the quantum spin-1/2 XY
spin chain). Alternatively, one may resort to a fully numerical approach, 
within which one numerically diagonalizes  $H_{2K}$ written in real space, 
as we do when explicitly computing the current entering a specific site.

\noindent
We now modify the clean Hamiltonian of Eq.(\ref{mh.1}) by adding the disorder on top of the clean system.

\subsection{Disordered longer-range Kitaev chain}
\label{disor}

Disorder is  introduced in many body systems by randomizing one, or more than one, parameters
of the system Hamiltonian. Accordingly,  physically observable quantities are first computed at 
a  given realization of the disorder and the final results are eventually averaged  over all the realizations of the disorder. In topological systems,  the interplay between  (a moderate amount of) disorder and topology may enforce 
 the topological phase, as compared to the 
clean limit, while a large value of disorder eventually leads to the disappearance of 
the topological phase(s) \cite{Brouwer2011,Pientka2012,Pientka2013,Nava2017,Lieu2018,Cinnirella2024,Cinnirella2025}. 
In some cases, some specific generalizations of $\omega$ can be defined even in 
the presence of disorder as it is the case, for instance, of the   ``disorder averaged winding number'' (DAWN),
which  can 
be unambiguously introduced only in the case in which the disordered Hamiltonian anticommutes
with a pertinently defined chiral symmetry operator at any realization of the disorder \cite{Mondragon2014,Liu2022}. 
In general, however, it is not clear how to detect the topology in the presence of disorder. 
(For instance, an  alternative means to probe the combined effects of topology and disorder
on the system is discussed  in \cite{Nava2023}). 

The phase diagram of the 2LRK with quenched Anderson disorder has been originally 
studied  in Ref.\cite{Lieu2018} by means of the transfer matrix method and by eventually employing the 
``entanglement degeneracy criterion'' to validate the phase diagram of the disordered system. Here, 
we generalize the model of \cite{Lieu2018} to several different types of disorder, as outlined below. 
Importantly, while the transfer matrix method applies well to the case of Anderson disorder, it is, in 
general, practically inapplicable to generalizations, such as the ones we consider here, including 
the case of the open chain connected to external reservoirs. This motivates 
us to resort to a different mean to characterize the phase diagram of the system by looking at the interface
current pattern of the chain connected to the reservoir through a metallic lead, as we discuss in the 
following of the paper. 

Formally, the generic disordered 2LRK model is described  by the Hamiltonian

\bea
    H_{2K,d} &=& 2\sum_{n=1}^L g_n \chi_n^\dagger \chi_n \nonumber \\ 
    & & -\sum_{n=1}^{L-1} \lambda_{1,n} \{\chi^\dagger_n \chi_{n+1}+\chi^\dagger_n \chi^\dagger_{n+1}+ {\rm h.c.} \} 
    \nonumber \\ 
    & &-\sum_{n=1}^{L-2} \lambda_{2,n}\{\chi^\dagger_n \chi_{n+2}+\chi^\dagger_n \chi^\dagger_{n+2}+ {\rm h.c.} \} 
    \:\: , 
    \label{eq:R2K_ham_disordered}
\eea
\noindent
with $\{g_n\},\{\lambda_{1,n} \},\{ \lambda_{2,n}\}$ being randomly distributed variables.  Specifically, in 
 the following we consider three different types of disorder, that is:
  
{\bf Type 1}: To validate our approach, we study the same system considered in \cite{Lieu2018}, that is, we introduce Anderson disorder by 
taking $\{\lambda_{1,n}\}$ and $\{\lambda_{2,n}\}$ to be uniform and not disordered, while we set 
 $g_n = \bar{g} + \epsilon_n$ with  $\bar{g}=1$ and the fluctuating random contributions $\{\epsilon_n\}$ to be distributed 
 following a uniform distribution such that $\epsilon_n \in [-\frac{\sqrt{3}W}{2},\frac{\sqrt{3}W}{2}]$,
 with $W$ being the strength of the disorder.

{\bf Type 2}: As a first example of correlated disorder in the chemical potential, we consider an Aubry-Andr\'e-Harper model 
\cite{Harper1955,Aubry1980}, with $g_n$ realized  by the quasi-periodic function $g_n = \bar{g} + \frac{W}{2} \cos{(2\pi \beta n + \phi)}$ with $\bar{g}=1$, 
$W$ being the disorder strength, $\beta=\frac{\sqrt{5}-1}{2}$ being the disorder period (incommensurate to the lattice period), 
and $\phi \in [0, 2\pi )$ being a random phase.

{\bf Type 3}:  Finally, we consider a correlated bond disorder (mimicking for example a shrinking or expansion of a link connecting two nearest neighboring sites and/or the presence of random magnetic fluxes \cite{Tadjine2018,Liu2022}). The disorder affects all the hoppings and the pairing strengths connected to the link. For each link connecting sites $n$ and $n+1$
the disorder acts on $\lambda_{1,n}$, $\lambda_{2,n}$ and $\lambda_{2,n-1}$.  In particular, 
we associate a probability $p_1$ to the case that the link is not affected by disorder, a probability $p_2$ to the case 
that the hopping and pairing strengths associated to the link increase, and a probability $p_3=1-p_1-p_2$ that the hopping and pairing strengths decrease (in the following we consider $p_1=p_2=p_3=1/3$). 
When the strengths increases(decreases), we set $\lambda_{1,n} = \lambda_1 \pm W$, $\lambda_{2,n} = \lambda_2 \pm W$,
and $\lambda_{2,n-1} = \lambda_1 \pm W$, with $W$ being the disorder strength and $\lambda_1$ and $\lambda_2$
being  the  values of the next- and second next-neighbor hopping and pairing strengths respectively, in the clean limit. 

\begin{figure}
\centering
\includegraphics[width=1.0\linewidth]{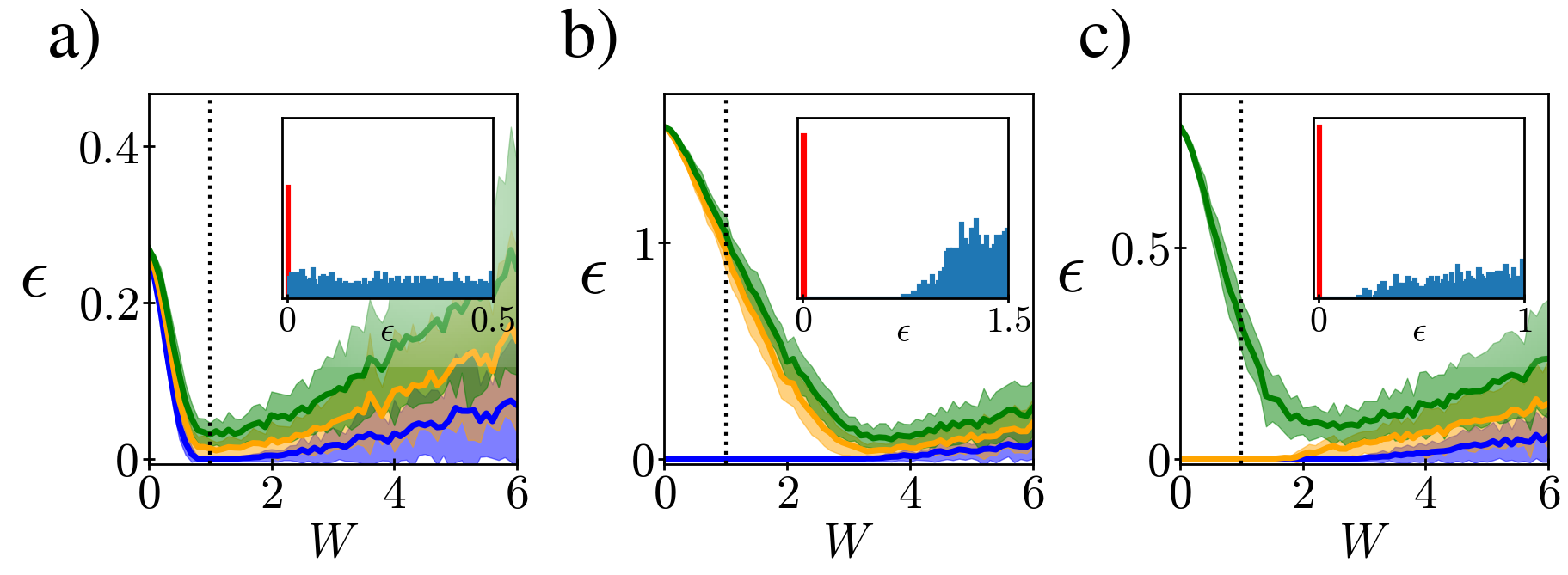}
\caption{Plot of the lowest three disorder averaged energy eigenvalues as a function of the Anderson disorder strength $W$
 for the longer-range Kitaev chain   with $g=1$, $L=100$ sites,  and
$\lambda_1=1.25,\lambda_2=-0.75$ (trivial phase - panel {\bf a)}), 
$\lambda_1=1.5,\lambda_2=0.5$ (topological  phase with $\omega=1$ - panel {\bf b)}),
 $\lambda_1=0.5,\lambda_2=-1.5$ (topological phase with $\omega = 2$ - panel {\bf c)}) and 50 disorder realizations. The 
 solid lines represent the average energy of the low-lying level (blue), of the next to low lying level (orange),
 and of the next to the next to the low lying level (green). The shaded area around the solid line measures
 the standard deviation of the energy eigenvalues around the average value. \\
 Inset (in all three cases): averaged level distribution computed at the value of $W$ corresponding to the
 dashed, vertical line. }
\label{disle.1}
\end{figure}

\begin{figure}
\centering
\includegraphics[width=1.0\linewidth]{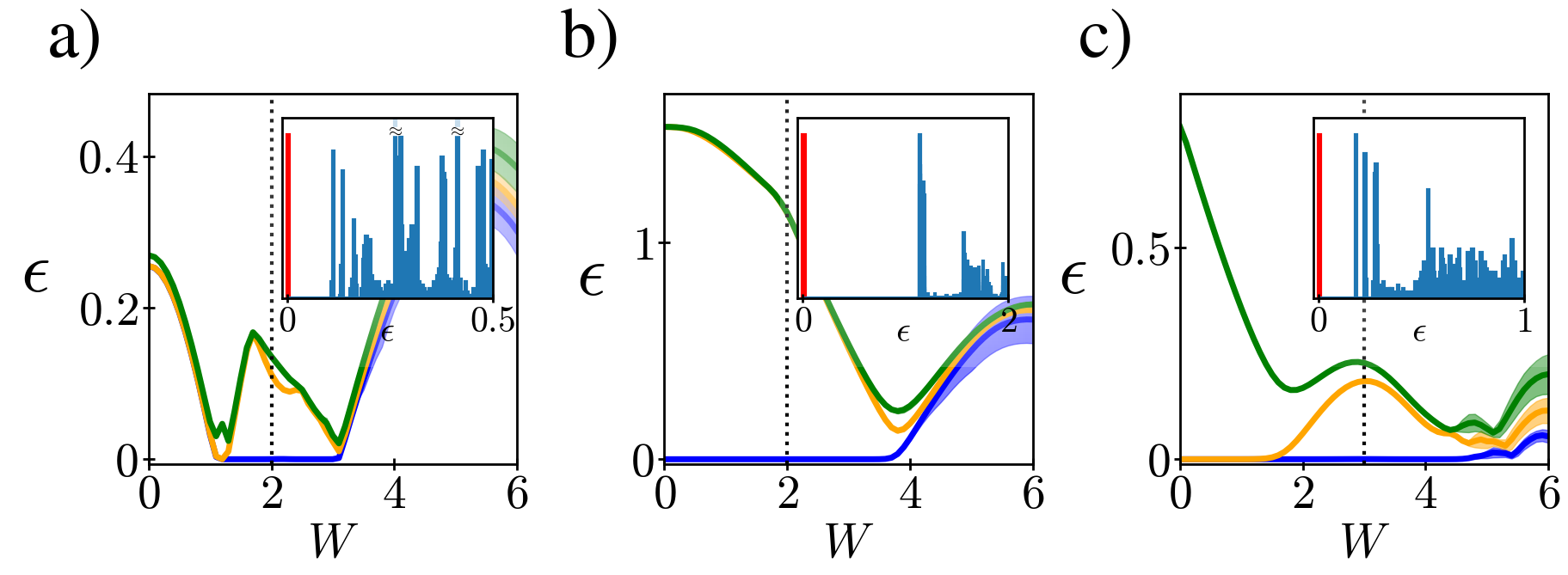}
\caption{Plot of the lowest three disorder averaged energy eigenvalues as a function of the AAH disorder strength $W$
 for the longer-range Kitaev chain   with $g=1$, $L=100$ sites,  and
$\lambda_1=1.25,\lambda_2=-0.75$ (trivial phase - panel {\bf a)}), 
$\lambda_1=1.5,\lambda_2=0.5$ (topological  phase with $\omega=1$ - panel {\bf b)}),
 $\lambda_1=0.5,\lambda_2=-1.5$ (topological phase with $\omega = 2$ - panel {\bf c)}) and 50 disorder realizations. The 
 solid lines represent the average energy of the low-lying level (blue), of the next to low lying level (orange),
 and of the next to the next to the low lying level (green). The shaded area around the solid line measures
 the standard deviation of the energy eigenvalues around the average value. \\
 Inset (in all three cases): averaged level distribution computed at the value of $W$ corresponding to the
 dashed, vertical line. }
\label{disle.2}
\end{figure}

\begin{figure}
\centering
\includegraphics[width=1.0\linewidth]{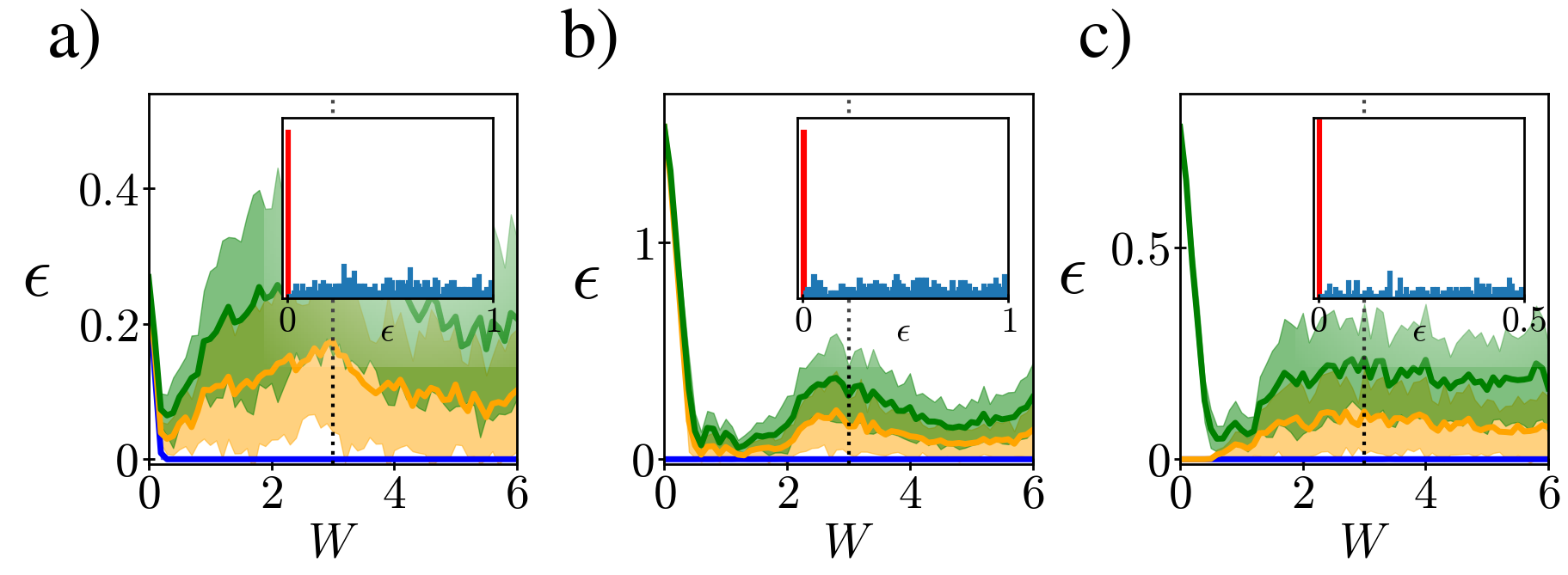}
\caption{Plot of the lowest three disorder averaged energy eigenvalues as a function of the correlated bond disorder strength $W$
 for the longer-range Kitaev chain   with $g=1$, $L=100$ sites,  and
$\lambda_1=1.25,\lambda_2=-0.75$ (trivial phase - panel {\bf a)}), 
$\lambda_1=1.5,\lambda_2=0.5$ (topological  phase with $\omega=1$ - panel {\bf b)}),
 $\lambda_1=0.5,\lambda_2=-1.5$ (topological phase with $\omega = 2$ - panel {\bf c)}) and 50 disorder realizations. The 
 solid lines represent the average energy of the low-lying level (blue), of the next to low lying level (orange),
 and of the next to the next to the low lying level (green). The shaded area around the solid line measures
 the standard deviation of the energy eigenvalues around the average value. \\
 Inset (in all three cases): averaged level distribution computed at the value of $W$ corresponding to the
 dashed, vertical line. }
\label{disle.3}
\end{figure}

To spell out the effects of the various types of disorder on the energy levels of the 
2LRK, we now proceed by   generating, for each kind of disorder,  a given number of configurations of 
the disorder, and by computing the energy spectrum for each disorder configuration.  Eventually, we 
  average the spectrum over all the disorder configurations. On going through that procedure
within a given window of the disorder strength $W$, we eventually draw plots of the low-lying energy levels
as a function of $W$. In Figs.\ref{disle.1},\ref{disle.2},\ref{disle.3} we show the corresponding results 
for the first three energy eigenvalues
in the case, respectively, of Anderson disorder, of the Aubry-Andr\'e-Harper model, and of the 
correlated bond disorder. In any one of the three cases we have drawn the plots for the chain with $g=1$, $L=100$ sites,  and
$\lambda_1=1.25,\lambda_2=-0.75$ (trivial phase - panel {\bf a)}), 
$\lambda_1=1.5,\lambda_2=0.5$ (topological  phase with $\omega=1$ - panel {\bf b)}),
 $\lambda_1=0.5,\lambda_2=-1.5$ (topological phase with $\omega = 2$ - panel {\bf c)}). The three points are marked as $A$, $B$ and $C$ in Fig.\ref{pha_dia}, respectively. Since, as an
 effect of the energy spreading due to the disorder, we have a distribution of energy eigenvalues
 for each level,  in all the figures 
 we followed the drawing code used in Ref.\cite{Cinnirella2025} and 
 depict with a solid line the  average energy of each level (blue for the lowest energy level, orange for 
 the next to lowest energy level, and green for the next to the next to lowest energy level), over 50 disorder realizations, with the shaded area around 
 each solid line representing the energy eigenvalue dispersion. In the trivial phase (panels {\bf a)}) there are no 
 real fermionic states at zero energy. As $W=0$ all three the low-lying states are located at energies larger than the 
 superconducting gap. Remarkably, on increasing $W$, the lowest-energy state, at some point, is pulled down to 
 zero energy, thus featuring what one would expect in the $\omega=1$ topological phase. Such a tendency to 
 generate a ``reentrant topological phase'' is known to take place, for instance, in the short-range Kitaev model
 in the presence of Anderson disorder \cite{Pientka2012,Pientka2013,Nava2017}. Here, we evidence how the
 feature is general with respect to both the 
 range of electron hopping and/or pairing in the Kitaev model and the specific kind of disorder one is considering. 
Panels {\bf b)} and  {\bf c)}, drawn for the system within a topological phase, show a similar trend, for a limited
amount of disorder, to generate a reentrant topological phase.

\noindent
 
\noindent
 
\noindent
To further evidence the revival/enforcement of the topological phases at a limited $W$, in the 
inset of all panels we show the full spectrum of the chain, averaged over 50 disorder realizations, 
taken at the value of $W$ corresponding to the vertical, dashed line in the main plots. In all three cases we clearly identify
a sharp peak pinned at zero energy, which corresponds to the localized real fermionic zero mode
that marks the topological phase. 

Comparing with each other the plots drawn for the same values of the system parameters, but for different 
kinds of disorder, we note that the energy broadening for the Anderson and for the correlated bond disorder is 
of the same order of magnitude, at any values of the parameters. Instead, in the AAH case we see almost
no level broadening, which is consistent with fact that in this case it is just the discommensurability effects
that mimic a disorder-induced randomness. Finally, we note that, in all three the cases,  at different
values of $W$ and over different energy scales, all the levels are fully gapped at a large enough disorder strength, 
which is a signal of the fact that a strong amount of disorder fully washes out nontrivial topological phases. 

In the following, we  extend our formalism to describe the dynamics of the open chain, connected 
via a metallic lead to the external reservoir, by means of the LME approach.

\subsection{Formal description of the chain connected to the external reservoirs via a metallic lead: review of 
Lindblad Master Equation approach}
\label{LME}

To induce electric current in the 2LRK, we connect one of its sites  to an external metallic lead, 
 so that, in its full generality, the Hamiltonian of the resulting system   reads $H = H_L+ H_{2K}+ H_T$, with:

\begin{eqnarray}
H_L &=& - t \:\sum_{i=1}^{{\cal N}-1} \{\alpha_i^\dagger \alpha_{i+1}
+{\rm h.c.} \} \nonumber \\ 
H_T &=& - \gamma  \{ \alpha_{{\cal N}}^\dagger \chi_n + {\rm h.c.} \}    
\:\: . 
\label{lme.1}
\end{eqnarray}
\noindent
In Eq.(\ref{lme.1}) we formally describe the metallic lead as a one-dimensional, spinless, noninteracting lattice electron
model at zero chemical potential and with hopping strength $t$, over an ${\cal N}$-site lattice \cite{Schwarz2016,Lotem2020,Leumer2021}. 
Accordingly, we   denote with $\alpha_i,\alpha_i^\dagger$ the
single-electron annihilation and creation operators at site-$i$, so that we get $\{\alpha_i,\alpha_{i'}^\dagger \} = 
 \delta_{i,i'}$, with all the other anticommutators being equal to 0. $H_T$ corresponds
to the single-electron tunneling operator between the endpoint of the lead and site-$n$ of the superconducting 
chain,   with corresponding tunneling strength   equal to  $\gamma$.  The lead  and the   superconducting
region constitute an hybrid system which, following the approach developed in
\cite{Guimaraes2016},  we now couple to the external reservoir  only through the  lead. Specifically, we regard the 
reservoir  as a source/sink of particles entering/exiting the lead at a given rate, which we specify in the following. In
this respect, the whole NS system works as an open system and, in addition, by keeping at a finite voltage bias  the
reservoir  coupled to the   lead, we inject/extract a finite current into the superconducting chain, through the metallic lead  
(note that, since
the key quantity we look at here is the dc conductance between the lead and the Kitaev chain, 
differently from what is done in \cite{Dabbuzzo2021}, we indirectly couple the Lindblad
bath to the superconductor via the metallic lead). 
  To describe the dynamics of our whole system, we resort to LME approach
\cite{Lindblad1976}. Specifically, we write the Schr\"odinger equation  for the density matrix of the system, $\rho (t)$, 
in the form 

\beq
\frac{d \rho (t)}{d t }   = - i [ H,\rho (t) ] + {\cal D} [\{\rho (t) \} ]
\;\; .
\label{lme.2}
\eneq
\noindent
In Eq.(\ref{lme.2}), $H$ is the total system Hamiltonian, while ${\cal D} [\{\rho (t) \}]$ is 
the so-called Lindbladian, which we are going to define next.

As a first step, we separately consider the Hamiltonian for the disconnected N and S regions, that is, 
we pretend that $\gamma = 0$, and diagonalize the two  of them separately. 
Denoting with $\eta_{L,k}$  the 
eigenmodes of the lead, we get 

\beq
\eta_{L,k}   = \sum_{j=1}^{\cal N} \: \sqrt{\frac{2}{{\cal N} + 1} } \: \sin (k j ) \: \alpha_j 
\:\: , 
\label{lme.3}
\eneq
\noindent
with $k=\frac{\pi \nu}{{\cal N} + 1}$, and $\nu = 1 , \ldots , {\cal N}$ and the corresponding 
single-particle energy $\epsilon_k = - 2 t \cos (k)$.  Next, we choose 
the  Lindblad jump operators,   so to 
  stabilize a NESS  in which particles injected/extracted 
through the leads are distributed according to Fermi distribution function. Following 
Refs.\cite{Nava2023,Cinnirella2024,Cinnirella2025,Lotem2020}, we construct the jump operators for the
 lead, $L_{( k,{\rm in}/{\rm out})}$, so that 

\begin{eqnarray}
L_{(L,k,{\rm in})} &=& \sqrt{f (\epsilon_k , V  , T  )} \eta_{L,k}^\dagger \nonumber \\
L_{(L,k,{\rm out})} &=& \sqrt{[1-f(\epsilon_k,V,T)]} \eta_{L,k} 
\:\: , 
\label{lme.3X}
\end{eqnarray}
\noindent
with the Fermi distribution for the  reservoir $f (\epsilon_k , V , T ) = 
\left\{1+ \exp \left[ \frac{\epsilon_k - V}{T} \right] \right\}^{-1}$ and $V$ being the bias of the reservoir connected to the lead. 
Given Eqs.(\ref{lme.3X}),  we 
define ${\cal D} [\{ \rho \}]$ as 

\beq
{\cal D} [\{ \rho (t)  \} ] = 2 \Gamma \sum_{\lambda } \left( L_\lambda \rho (t) L_\lambda^\dagger 
- \frac{1}{2}   \{L_\lambda^\dagger L_\lambda , \rho (t) \}   \right) 
\;\; , 
\label{lme.4}
\eneq
\noindent
with $\lambda \in \{ (k,{\rm in}),(k,{\rm out})\}$. Also, in 
Eq.(\ref{lme.4}) we have chosen the coupling $\Gamma$ between the leads and the reservoirs to be
independent of the quantum numbers associated to the eigenmodes, so to avoid unnecessary 
complications in the following derivation.  Similarly to what we do in Eqs.(\ref{lme.3}), we introduce 
the eigenmodes of the open, isolated Kitaev chain, $\eta_{2K,q}$, by means of the Bogoliubov-Valatin 
transformations 

\begin{eqnarray}
\eta_{2K,q} &=& \sum_{n=1}^{L} \{ u_{q,n} \chi_n + v_{q,n} \chi_n^\dagger \}   \nonumber \\
\eta_{2K,q}^\dagger &=& \sum_{n=1}^L \{ [v_{q,n}]^* \chi_n + [u_{q,n}]^* \chi_n^\dagger \} 
\:\: . 
\label{lme.5}
\end{eqnarray}
\noindent
Using  Eqs.(\ref{lme.3},\ref{lme.5}), we may rewrite the Hamiltonian $H$ as 

\beq
H = \sum_{\alpha,\beta}   \{ \eta_\alpha^\dagger  A_{\alpha,\beta} \: \eta_\beta  + \eta_\alpha^\dagger B_{\alpha,\beta}  \: \eta_\beta^\dagger 
+ \eta_\alpha B^*_{\beta,\alpha}   \: \eta_\beta \} 
\;\; , 
\label{lme.6}
\eneq
\noindent
with $\alpha,\beta \in \{(L,k),(2K,q)\}$ and the matrices $A$ and $B$ defined from 
the Hamiltonian expressed in terms of the whole set of the $\eta$-operators. 
Importantly, the presence of the superconducting block, corresponding to the 
longer-range Kitaev chain, yields the anomalous pairing terms at the right hand side
of Eq.(\ref{lme.6}), encoded in the nonzero $B$ matrix. 

Using LME approach reviewed above, in the following we compute the current exchanged between
the lead and site-$n$ of the superconducting chain, which we use to map out the phase diagram of the longer-range disordered 
Kitaev chain. To do so, the key quantity we need  is the 
correlation matrix $\theta (t)$, together with its anomalous counterpart, $\theta^A (t)$. Their 
matrix elements are respectively defined as 

\begin{eqnarray}
\theta_{\alpha,\beta}   (t) &=& {\rm Tr} \{\eta_\alpha^\dagger \eta_\beta \rho (t) \}   \nonumber \\
\theta^A_{\alpha,\beta} (t) &=& {\rm Tr} \{\eta_\alpha^\dagger \eta_\beta^\dagger \rho (t) \}
\:\: . 
\label{lme.7}
\end{eqnarray}
\noindent
Starting from  Eq.(\ref{lme.2}), we now  write the 
time evolution equations for $\theta(t)$ and for $\theta^A (t)$. In doing so, the crucial point
is that the set of equations for the corresponding matrix elements is closed, that is, no 
time-dependent correlation functions of more than two single-particle operators are involved.
As a result, we obtain \cite{Cinnirella2025} 

\begin{eqnarray}
\frac{d \theta (t)}{d t}   &=& i [A^T,\theta(t)]   + 2 i (\theta^A (t) B + B^*[\theta^A (t)]^\dagger ) \nonumber \\ & &
- \frac{1}{2}  \{G + R , \theta (t) \}  + G \:\: , \nonumber \\
\frac{d \theta^A (t)}{dt} &=& i (\theta^A (t) A + A^T \theta^A (t)) - 2 i (B^*\theta^T (t) + \theta (t) B^* )  \nonumber \\ & &
-\frac{1}{2}   \{G+R,\theta^A (t) \} + 2 i B^* 
\:\: ,
\label{lme.8}
\end{eqnarray}
\noindent
with $^T$ denoting the matrix transpose.
We have already defined the matrices $A$ and $B$ in Eqs.(\ref{lme.6}). 
The $G$ and $R$ matrix elements are, instead, defined as 

\begin{eqnarray}
G_{\alpha,\beta}   &=& 2  \Gamma    \delta_{\alpha,\beta}   f_\alpha \nonumber \\
R_{\alpha,\beta} &=& 2  \Gamma    \delta_{\alpha,\beta}   \{1-f_\alpha \} 
\;\; , 
\label{lme.9}
\end{eqnarray}
\noindent
with 

\beq
f_\alpha = \Biggl\{ \begin{array}{l} f (\epsilon_k,V,T) \;\; {\rm if} \: \alpha = (L,k)  \\
0 \;\; {\rm if}   \: \alpha = (2K,q) 
\end{array}
\:\: .
\label{lme.10}
\eneq
\noindent
In the following, we employ Eqs.(\ref{lme.8}) to compute the current across the NS interfaces in 
the non-equilibrium steady state of our system.

\section{DC conductance in the non-equilibrium steady state and phase diagram of the disordered
system}
\label{cness}

The dc current at the NS interface between the endpoint of the normal lead and the $n$-th site of the
 longer-range Kitaev chain is given by (see Eq.(\ref{ap.2.1}) for the definition of the corresponding operator)

\beq
I_{{\rm int},n} = - i \gamma \{\langle \alpha_{{\cal N}}^\dagger \chi_n \rangle - \langle 
\chi_n^\dagger \alpha_{{\cal N}} \rangle \} 
\:\: , 
\label{res.1}
\eneq
\noindent
with $\langle  \ldots \rangle$ denoting the expectation value of the operator computed in the system at  the NESS.
In Appendix \ref{current_o} we discuss in detail how to compute the average values of the bilinear of fermionic 
operators at the right-hand side of Eq.(\ref{res.1}) in terms of the correlation matrices at the NESS, 
$\theta_{\rm NESS}$ and $\theta^A_{\rm NESS}$.  The key point in using the quantity in 
Eq.(\ref{res.1}) to compute the topological invariant of the system is that, as we  prove in Appendix  \ref{current_o},
 the system itself lies within a gapped phase. Within linear response theory in the voltage bias $V$ between
the metallic lead and the superconducting chain, one obtains

\beq
I_{\rm int} = \sum_{n =1}^L I_{{\rm int},n} = G_{\rm int}   V 
\;\; , 
\label{res.x1}
\eneq
\noindent
with the total dc conductance being given by 

\beq
G_{\rm int} =  
\frac{2 \gamma^2}{\pi t \Gamma} \left[1 + \left(\frac{\Gamma}{2t} \right)^2 \right] \omega 
\:\: . 
\label{res.x2}
\eneq
\noindent
In Eq.(\ref{lme.1}), the parameters $t$ and $\gamma$ respectively correspond to the single-fermion 
hopping strength between nearest neighboring sites of the metallic leads and between the endpoint of 
the lead and the site $n$ of the 2LRK. $\Gamma$ is introduced in Eq.(\ref{lme.4}) and 
corresponds to the coupling strength between the superconducting chain and the Lindblad bath.  $\omega$ is the 
winding number, basically counting the number of localized, subgap states that set in within the topological phase.
As a result, we find that, by fixing the system parameters and measuring $G_{\rm int}$ in units of 
$G_{\rm int}^{(0)} = \frac{2 \gamma^2}{\pi t \Gamma} \left[1 + \left(\frac{\Gamma}{2t} \right)^2 \right] $, 
one directly gets 

\beq
\frac{ G_{\rm int} }{ G_{\rm int}^{(0)} }=  \omega
\:\: . 
\label{res.x3}
\eneq
\noindent
Eq.(\ref{res.x3}) is the key relation behind the derivation of this Section, where we employ the result 
of the numerical calculation of the dc conductances, in the clean limit as well as in the presence of 
disorder, to map out the whole phase diagram of our system. 
Indeed, Eq.(\ref{res.x3}) is expected to keep valid in the presence of disorder, as long as the energy of the 
subgap states induced by disorder is greater than the applied bias $V$. However, as we will show in the following, 
also for strong disorder (at least for $W\approx 3 \times  \rm{max}(g,\lambda_1,\lambda_2)$), the (disorder 
averaged) $\omega$ proves to be a robust probe of the topological phase, due to the fact that the non-quantized contribution
 to the conductance due the ``bulk'', over the gap,  states (Eq.(\ref{ap.2.3}) of Appendix \ref{current_o}), is much smaller 
 than the topological one (Eq.(\ref{ap.2.4})), proportional to $\omega$. Furthermore, in order to probe $\omega$ 
 for greater values of $W$ and/or for non-zero temperature, $T$, one can extract the zero-bias conductance
  with higher precision (thus washing out the contribution of the disorder induced subgap state) by increasing 
  the length of the normal lead, and thus having access to lower values of $V$, or by implementing a 
  logarithmic-linear discretization scheme for the normal lead density of states \cite{Schwarz2016,Lotem2020}.
  
  In the following we numerically compute $\omega$ setting $\mathcal{N}=L=50$, $g=1$, $t=0.05$, $\gamma=0.01$ and $\Gamma=0.005$. Although working in the small-$V$ limit, we carefully take $V$ larger than the finite-size gap of 
  the metallic lead, so to avoid spoiling of our results by finite-size effects.    Finally, we fix   $\lambda_1=0.5$ and investigate the phase diagram
   of the disordered 2LRK  as a function of the next-to-nearest single electron hopping and pairing strengths 
  $\lambda_2$ and of the disorder strength $W$ for the three types of disorder introduced in Sec.\ref{disor}.

\begin{figure}
\includegraphics[width=0.9\linewidth]{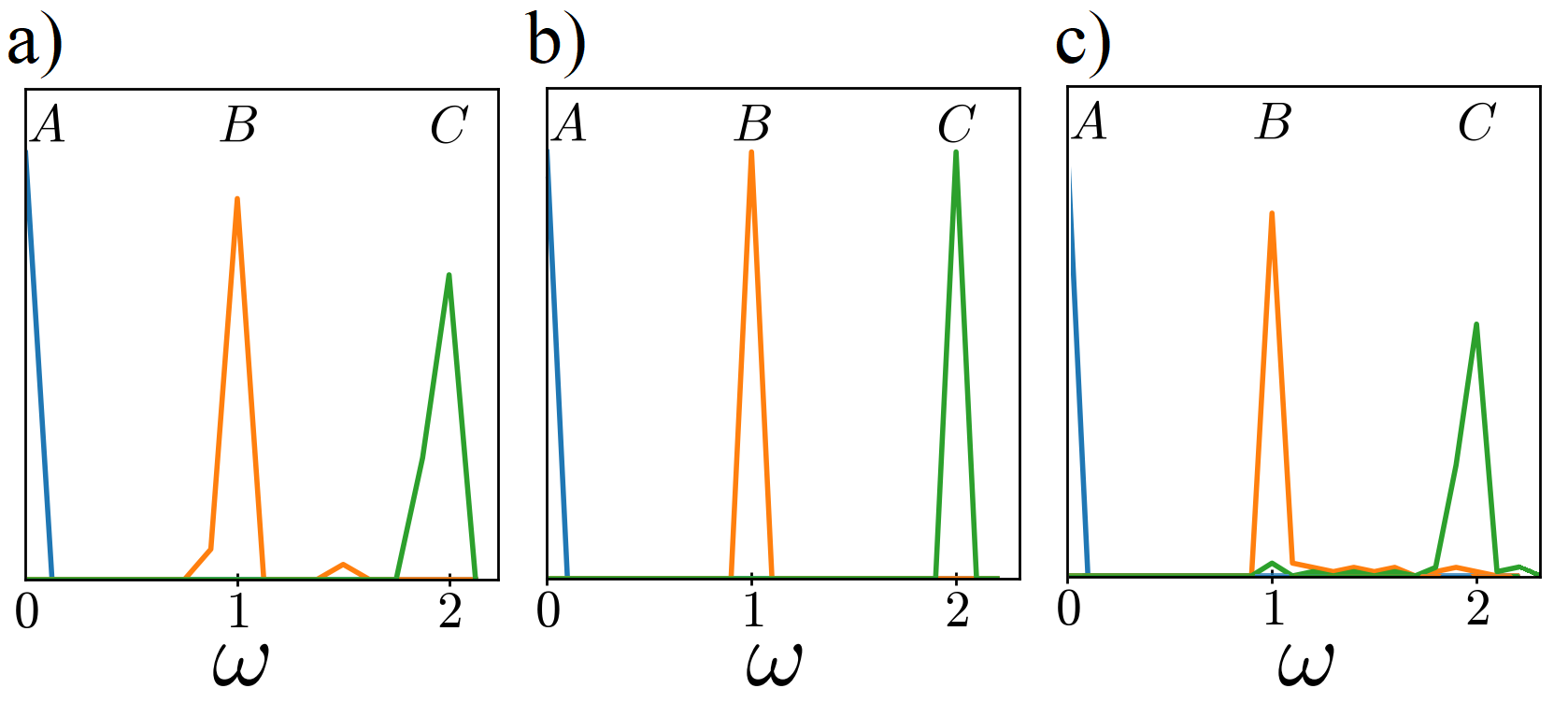}
\caption{Histograms of $\omega$, extracted from the dc conductance in Eq.(\ref{res.x2}) between  a normal lead 
connected to a longer-range Kitaev chain and the chain itself, in the presence of different types of disorder (see Sec.\ref{disor}). 
In all panels we set $\mathcal{N}=L=50$, $g=1$, $t=0.05$, $\gamma=0.01$, $\Gamma=0.005$ and $\lambda_1=0.5$. 
Points $A$, $B$ and $C$ correspond to different values of $\lambda_2$ and $W$. Specifically, we consider the following 
kinds of disorder: \\
Panel {\bf a)}: Anderson disorder (Type 1) for $\lambda_{2,A}=-0.6$, $W_A=0.5$, $\lambda_{2,B}=0.8$, $W_B=1.5$,
 and $\lambda_{2,C}=-1.6$, $W_C=2.5$; \\
  Panel {\bf b)}: Aubry-Andr\'e-Harper correlated disorder (Type 2) for $\lambda_{2,A}=-0.8$, $W_A=0.3$, $\lambda_{2,B}=-0.8$, 
  $W_B=2.0$, and $\lambda_{2,C}=-1.8$, $W_C=2.0$; \\
   Panel {\bf c)}: correlated bond disorder (Type 3) for $\lambda_{2,A}=-0.8$, $W_A=0.3$, $\lambda_{2,B}=-0.8$, $W_B=1.0$ and $\lambda_{2,C}=-1.8$, $W_C=4.5$. 
 }
\label{histo}
\end{figure}
\noindent
To check that the quantization of $\omega$ holds also in presence of disorder, in Fig.\ref{histo} we plot an histogram of the 
values of $\omega$, extracted from the numerical results for  $G_{\rm int}$ computed  for 100 realizations of the disorder 
and for different combinations of $(\lambda_2,W)$.

In Fig.\ref{histo}a) we plot the distribution of values of $\omega$   for the Anderson disorder, with
points $A$, $B$ and $C$ corresponding to the following pairs of values of  $(\lambda_2,\: W)$:
$A=(-0.6,0.5)$, $B=(0.8,1.5)$ and $C=(-1.6,2.5)$. Case $A$ shows 
an histogram peaked at $\omega_A=0$, corresponding to a point of the phase diagram in a topologically trivial phase, 
hosting no topological states; the histogram for case $B$ is peaked around $\omega_B=1$, a signature of a topologically nontrivial 
phase hosting a single topological state at each boundary; finally, case $C$ histogram is peaked around $\omega_C=2$, 
associated to a topologically nontrivial phase with two topological states at each boundary. A similar result holds in Fig.\ref{histo}b), 
for the Aubry-Andr\'e-Harper disorder, 
where we have chosen $A=(-0.8,0.3)$, $B=(-0.8,2.0)$ and $C=(-1.8,2.0)$ and in Fig.\ref{histo}c), Type 3 disorder 
(correlated bond disorder), with $A=(-0.8,0.3)$, $B=(-0.8,1.0)$ and $C=(-1.8,4.5)$. 
It is worth to note that,   even for 
 high values of $W$, the quantization of $\omega$ is preserved. 

\begin{figure}
\includegraphics[width=1.0\linewidth]{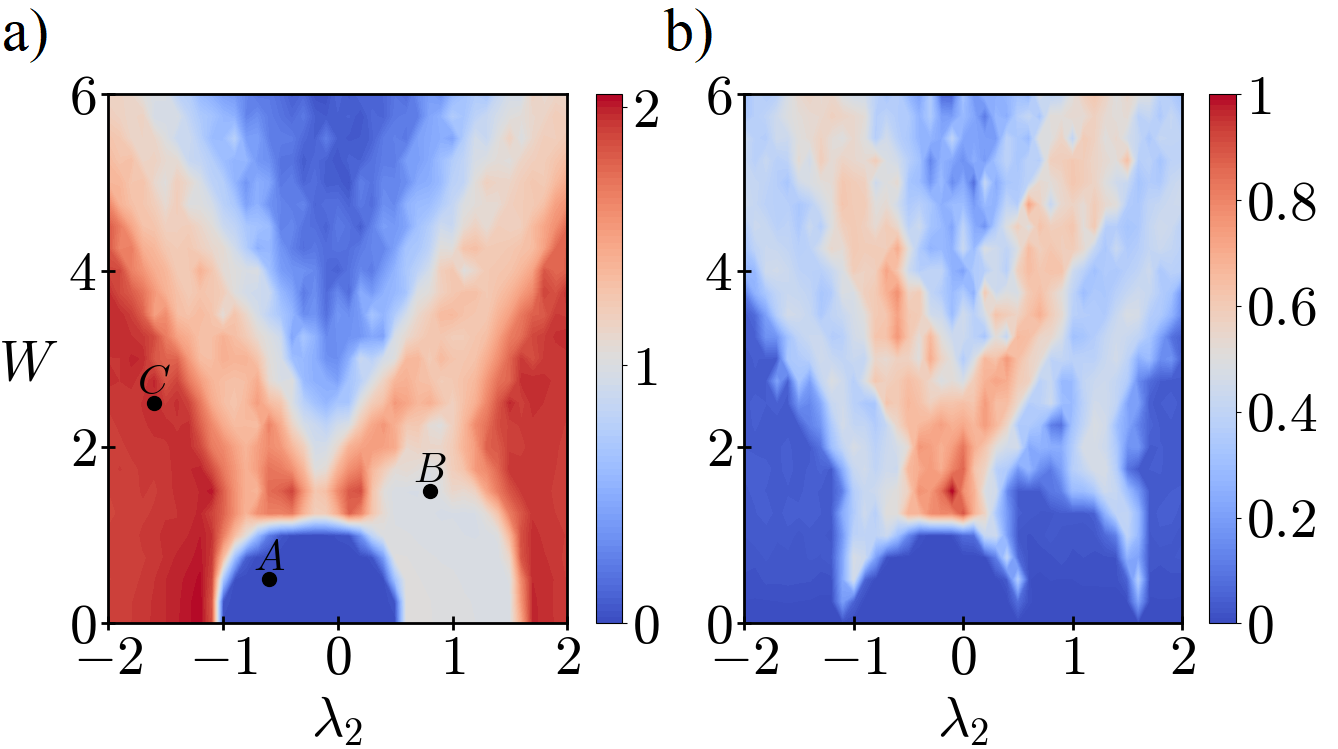}
\caption{Phase diagram of the longer-range Kitaev model of length $L=50$ and Type 1 disorder in the $\lambda_2-W$ plane. The other parameters are the same as in Fig.\ref{histo}. Panel {\bf a)} $\bar{\omega}$  averaged over 100 disorder realizations. Points $A$, $B$ and $C$ are the same as listed in Fig.\ref{histo}a). Panel {\bf b)} standard deviation of $\bar{\omega}$.
}
\label{and_phase}
\end{figure}
\noindent
In Fig.\ref{and_phase} we reconstruct the phase diagram in the $\lambda_2-W$ plane for the Type 1 disorder. In panel {\bf a)} we 
show the mean value of $\omega$ over 100 disorder realization, $\bar{\omega}$, while in panel {\bf b)} the provide corresponding standard deviation.  (We also
highlight the points $A$, $B$ and $C$ shown in Fig.\ref{histo}a)) The phase diagram exhibits bulk areas of uniform color and low values of the standard 
deviation. In these regions $\bar{\omega}$ is equal to $0$ (no topological states), $1$ (one topological state at each boundary) 
and $2$ (two topological states for boundary). These bulk areas are separated by extended regions characterized by intermediate 
values of $\bar{\omega}$ and high values of the standard deviation, where the topological phase transitions take place. 
Along the phase transitions, despite a non quantized value of $\bar{\omega}$, for each disorder realization we have a well
 quantized value of $\omega$. The histograms of $\omega$ in these regions show a bimodal distribution peaked around 
 integer values of $\omega$ with different reciprocal altitudes of the peaks, similar to the Griffiths phase observed in many topological
  models and by means of several topological invariants \cite{Motrunich2001, Sau2012,Nava2017,Nava2023,Cinnirella2025}. 
Focusing on Fig.\ref{and_phase}a) some interesting features emerge. At weak disorder, around $\lambda_2=0$, the topological 
trivial phase, blue region, characterized by $\bar{\omega}=0$, fades away and is replaced by the grey and the red regions. 
This is a signature of a disorder induced reentrant topological phase transition. in favor of phases hosting a single or two topological
 modes at each boundary \cite{Nava2017,Pientka2013,Zuo2022}. Further increasing disorder strength drives again the system toward
  a topological trivial phase, washing out the presence of the topological states. It is worth to note that the phase diagram derived in our
   paper for the Anderson disorder, by computing the interface current, agrees with the one analytically predicted in \cite{Lieu2018} by  transfer
matrix and entanglement metrics techniques. However, analytical approaches, like the one in \cite{Lieu2018} or, as well, the 
strong disorder renormalization group approach \cite{Ma1979,Dasgupta1980,Fisher1992}, typically provide sharp transitions lines,
 without the possibility of resolving the transition regions that separate different topological phases \cite{Cinnirella2024}.

\begin{figure}
\includegraphics[width=1.0\linewidth]{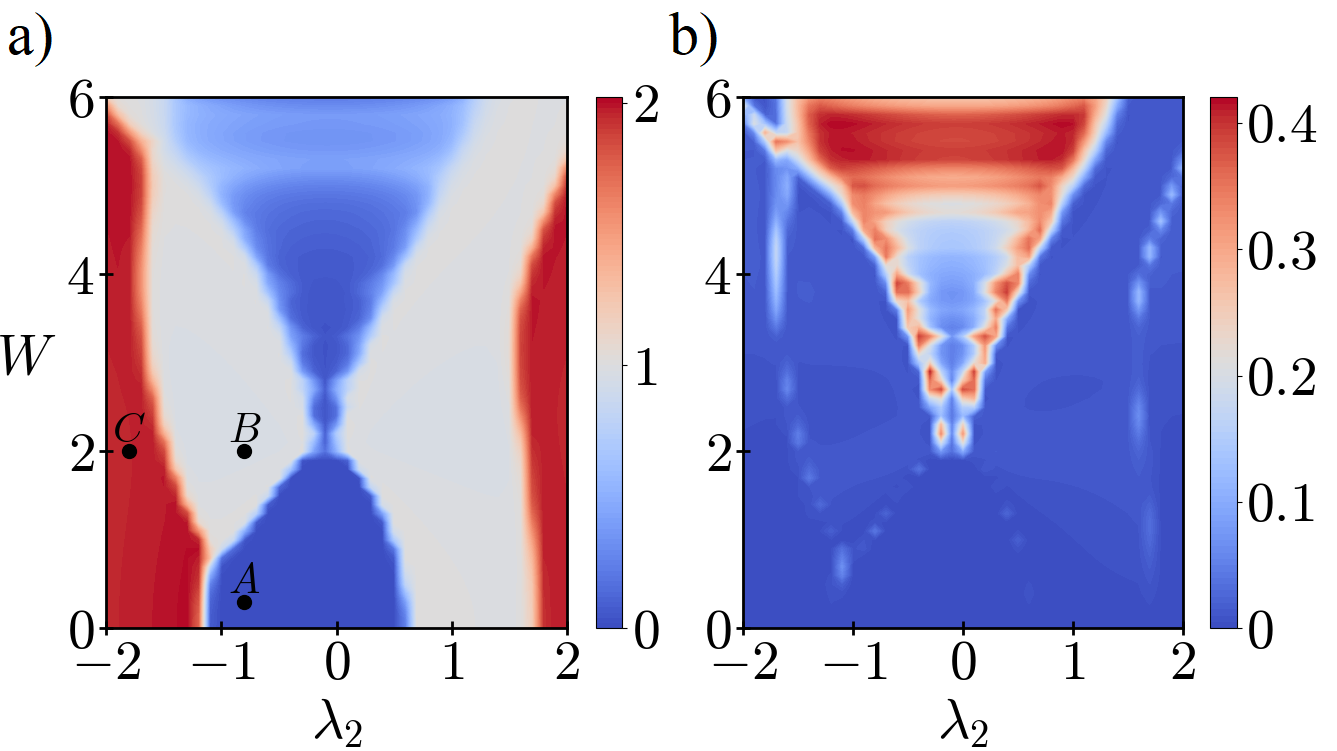}
\caption{Phase diagram of the longer-range Kitaev model of length $L=50$ and Type 2 disorder in the $\lambda_2-W$ plane. The other parameters are the same as in Fig.\ref{histo}. Panel {\bf a)} $\bar{\omega}$  averaged over 100 disorder realizations. Points $A$, $B$ and $C$ are the same as listed in Fig.\ref{histo}b). Panel {\bf b)} standard deviation of $\bar{\omega}$.
}
\label{AAH_phase}
\end{figure}
\noindent
In Fig.\ref{AAH_phase} we show $\bar{\omega}$ and its standard deviation in the $\lambda_2-W$ plane for the Type 2 disorder. 
The Aubry-Andr\'e-Harper disorder is a correlated disorder
characterized by some interesting properties. First, compared to Anderson disorder, in one-dimensional models
 it exhibits a localization transition for a finite 
value of disorder strength \cite{Dominguez2019,Berger2024}. Second, in long range topological models it can lead to the 
creation 
of disorder generated topological modes \cite{Fraxanet2021}. Both features can be observed in Fig.\ref{AAH_phase}a). 
Indeed, for $W<0.5$ there is barely no effect due to the disorder while we observe the onset of a reentrant topological phase for the phase with $\omega=1$ (grey area) only for $W>0.5$. This transition is similar to the one observed for Anderson disorder and is due the fact that a weak localization is known to protect the topological states \cite{Nava2017}. However, this property does not hold for the topological phase 
with $\bar{\omega}=2$ (red area) that retreats with increasing $W$. Starting from $W\approx 2$ the $\bar{\omega}=0$
 region begins to expand, as also observed for Type 1 disorder. Eventually, a strong enough disorder is expected to suppress 
 nontrivial topological phases. Compared with the Type 1 disorder, the boundary regions are now much sharper, as highlighted by a small standard deviation in almost the entire phase diagram, see Fig.\ref{AAH_phase}b). However, for 
 $W>5$, we observe a transition from a trivial phase towards a topological non trivial phase, together with a sharp increase of 
 the standard deviation. Contrarily to the reentrant topological phase transition observed for lower values of disorder, we 
 expect that this high disorder region is due the creation of disorder generated topological modes similar 
 to the ones observed in the infinite range Kitaev model \cite{Fraxanet2021}.

\begin{figure}
\includegraphics[width=1.0\linewidth]{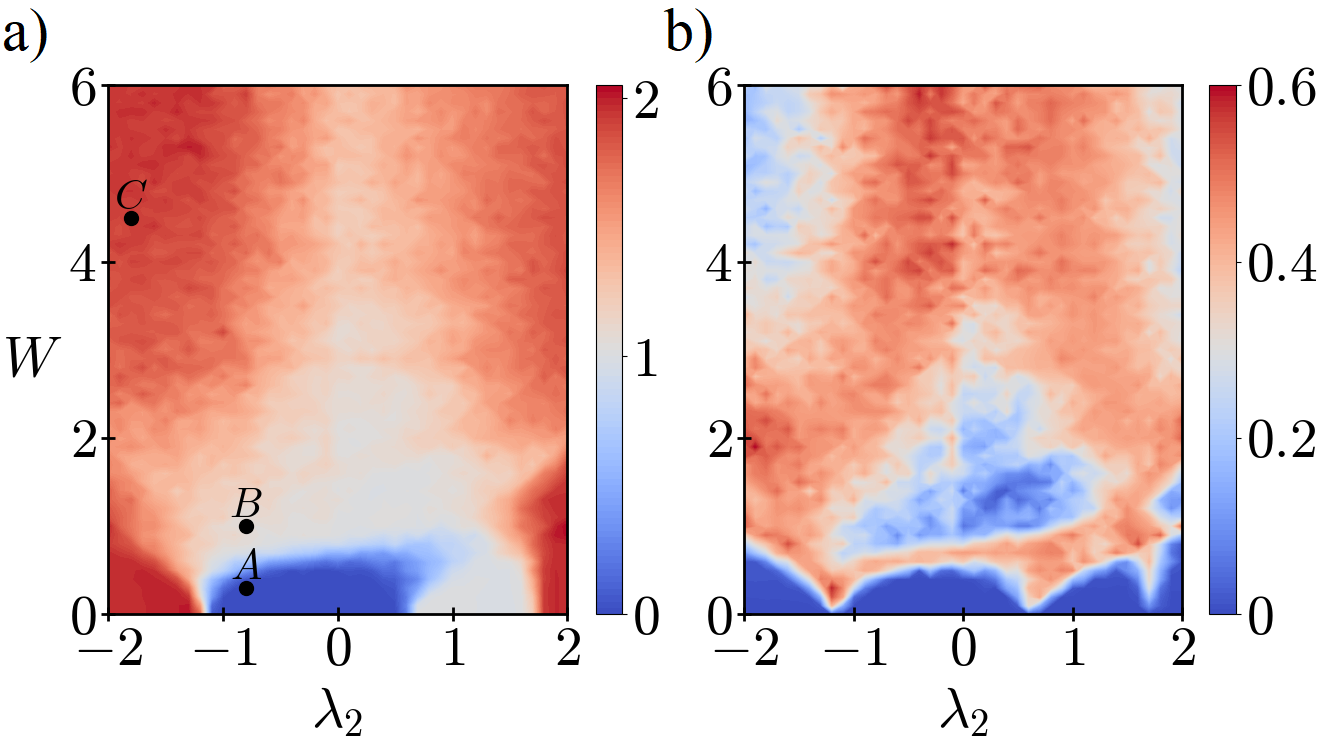}
\caption{Phase diagram of the longer-range Kitaev model of length $L=50$ and Type 3 disorder in the $\lambda_2-W$ plane. The other parameters are the same as in Fig.\ref{histo}. Panel {\bf a)} $\bar{\omega}$  averaged over 100 disorder realizations. Points $A$, $B$ and $C$ are the same as  in Fig.\ref{histo}c). Panel {\bf b)} standard deviation of $\bar{\omega}$.
}
\label{lambda_phase}
\end{figure}
\noindent

Finally, in Fig.\ref{lambda_phase} we show $\bar{\omega}$ and its standard deviation in the $\lambda_2-W$ plane for the correlated 
disorder of Type 3. At weak disorder strength this type of disorder quickly suppresses both the topological trivial phase (blue area) and
 the $\bar{\omega}=2$ topological one, in favor of the topological phase with $\bar{\omega}=1$. However, for strong disorder, 
 compared to the other types of disorder, the trivial phase is not recovered and, on the contrary we have a second 
 reentrant topological transition in favor of the $\bar{\omega}=2$ phase, with a drastic reduction of the standard deviation 
 around $\lambda_2=\pm 2$.
In conclusion, we have shown that, for non-correlated Anderson disorder (Type 1), a reentrant topological transition
 is possible in the low disorder region of the phase diagram, while stronger values of the disorder tend to wash out the
  topological states. On the contrary, correlated disorder (Type 2 and Type 3) can protect, and eventually generate, 
  the topological phases also at high values of the disorder strength.

\section{Conclusions}
\label{concl}

In this paper we compute the disorder averaged dc conductance  between a metallic lead connected to a site 
of a disordered 2LRK,  and the chain itself, in the NESS toward which the system
evolves on connecting the lead to an external reservoir. Varying the  parameters of the 2LRK Hamiltonian 
and the disorder strength, we use our results to map out the whole phase diagram of our system, for 
three different types of disorder. Doing so, we are able to
evidence how disorder affects the different topological phases that set in the 2LRK in the clean
limit, eventually showing how, while (a weak amount of) Anderson disorder enforces low-$|\omega|$ topological 
phases, while suppressing phases with higher winding number, the other two types of (correlated) disorder
further stabilize the higher-$|\omega|$ phases, as well, or allow for a reentrant topological phase at strong disorder. 

 Given its generality and its ease of implementation, our approach is 
amenable of straightforward generalizations to topological systems with a more complex phase diagram 
than the 2LRK and in more complex geometries \cite{Li2021,Nehra_2020,Guerci2021,Langari_2015,Zhou_2016},
 thus providing a general and quite effective method to unveil 
the intriguing interplay between disorder and topology 
in topological superconducting systems. 
  
\begin{acknowledgments}
A. N. acknowledges funding by the Deutsche Forschungsgemeinschaft (DFG, German Research Foundation),
Projektnummer 277101999 -- TRR 183 (project B02), under project No.~EG 96/13-1,
and under Germany's Excellence Strategy -- Cluster of Excellence Matter and Light for 
Quantum Computing (ML4Q) EXC 2004/1 -- 390534769.  
 E. G. C. acknowledges funding (partially) supported by ICSC
Centro Nazionale di Ricerca in High Performance Computing, Big Data and Quantum Computing, funded by European
Union NextGenerationEU and thanks the Quantum Matter Institute - Vancouver, for the kind hospitality during the completion of this work.
The data underlying the figures in this work can be 
found at the zenodo site: \url{https://doi.org/10.5281/zenodo.16947234}
\end{acknowledgments}

\appendix 

\section{Mathematical derivation of the dc conductance  through the interfaces between the Kitaev chain and the metallic 
leads}
\label{current_o}

In this Appendix we derive the formula for the dc conductance between the metallic lead and a site
of the 2LRK  and discuss its relation 
with the  (disorder averaged) $\omega$. To do so, we rely on the formalism
of Ref.\cite{Guimaraes2016}, although pertinently adapted to our problem. To make the presentation of 
our method the clearest it is possible, in the following we summarize the various formal steps in a sequence
of subsections. While, strictly speaking, the approach we present here
applies in the clean limit only, it provides an important guideline to the extension of our method to 
the disordered system, which we present in the main text. 

\subsection{Nambu representation and generalized Linblad equation for the correlation matrix elements}
\label{sub_cur_1}

To ease the following following formal steps, we define the ($2({\cal N} + L) \times 2({\cal N} + L)$) 
correlation matrix in Nambu basis, $\Theta (t)$, as

\beq
\Theta (t) = \left[\begin{array}{cc}   \theta (t) & \theta^A (t) \\
(\theta^A (t) )^\dagger  & {\bf I}-\theta^T (t)  \end{array}   \right]
\;\; , 
\label{ap.1.1}
\eneq 
\noindent
with ${\bf I}$ being the identity matrix. In terms of $\Theta (t)$, we can rewrite Eqs.(\ref{lme.8}) in 
a compact form as 

\beq
\frac{d \Theta (t)}{dt} = i [{\cal K} , \Theta (t)]- \frac{1}{2}   \{{\cal G} + {\cal R},\Theta (t) \}   + 
{\cal G}   
\;\; , 
\label{ap.1.2}
\eneq
\noindent
with 

\beq
{\cal K}   = \left[\begin{array}{cc}   A^T & - 2 B^\dagger \\
- 2 B & - A \end{array} \right]
\;\; , 
\label{ap.1.3}
\eneq
\noindent
and

\beq
{\cal G} = \left[  \begin{array}{cc} G & {\bf 0} \\ {\bf 0} & R \end{array} \right] 
\;\; , \; {\cal R} = \left[\begin{array}{cc} R & {\bf 0} \\ {\bf 0} & G \end{array} \right] 
\:\: . 
\label{ap.1.4}
\eneq
\noindent
Eq.(\ref{ap.1.2}) is the key ingredient required to derive the correlation matrix characterizing the non-equilbrium 
steady state, $\Theta_{\rm NESS}$, which has to be independent of time and, therefore, has to satisfy the equation 
$\frac{d \Theta_{\rm NESS} (t)}{dt} = 0$. To explicitly solve it for 
$\Theta_{\rm NESS}$, we write it in a series expansion of the couplings between the superconducting chain and 
the metallic lead, $\gamma$.
Specifically, we set 

\beq
\Theta_{\rm NESS} = \sum_{n=0}^\infty \: \gamma \Theta_{{\rm NESS},n}  
\;\; , 
\label{ap.1.5}
\eneq
\noindent
and, to rewrite the full condition for the non-equilibrium steady state as a set
of recursive equations for the $\Theta_n$'s, we split ${\cal K}$ as 

\beq
{\cal K} = {\cal W}_0 - \gamma {\cal W}_1 
\;\; , 
\label{ap.1.6}
\eneq
\noindent
with 

\beq
{\cal W}_0 = \left[\begin{array}{cc}   {\bf W}_0 & {\bf 0} \\ 
{\bf 0} & - {\bf W}_0 \end{array} \right]
\;\; , 
\label{ap.1.7}
\eneq
\noindent
and $[ {\bf W}_0 ]_{\alpha,\beta}   = \delta_{\alpha,\beta} \epsilon_\alpha$, with $\epsilon_\alpha$ being the corresponding energy
eigenvalue of the {\it disconnected}   system, in which one has set $\gamma = 0$. A sequence of interconnected
equations for the $\Theta_N$ is generated by singling out the coefficients of the terms at any order in $\gamma$, 
thus obtaining, at the NESS

\begin{eqnarray}
D [\Theta_{{\rm NESS},0}] &=& {\cal G} \nonumber \\
D [\Theta_{{\rm NESS},n} ] &=& -i [{\cal W}_1 , \Theta_{{\rm NESS},n-1}] 
\;\, , 
\label{ap.1.8}
\end{eqnarray}
\noindent
with $n \geq 1$.  The solution of Eqs.(\ref{ap.1.8}) can be readily provided in terms
of the matrix elements of $\Theta_{\rm NESS}$, $[\Theta_{\rm NESS}]_{\alpha,\beta}$, with 
$\alpha,\beta    \in \{(L,k,p),(2K,q,p)\}$ and $p = \pm 1$ being the Nambu index. As a result, 
one obtains 

\onecolumngrid
\begin{eqnarray}
[ \Theta_{{\rm NESS},0}]_{\alpha,\beta }   &=& 
\frac{{\cal G}_{\alpha,\beta}}{-i ([{\cal W}_0]_{\alpha,\alpha} + 
[{\cal W}_0]_{\beta,\beta}) +\frac{1}{2}  [{\cal G}   + {\cal R} ]_{\alpha,\alpha} + 
 [{\cal G}   + {\cal R} ]_{\beta,\beta}} \nonumber \\
 [\Theta_{{\rm NESS},n}]_{\alpha,\beta} &=&
  \frac{-i [{\cal W}_1, [\Theta_{{\rm NESS},n-1} ]]_{\alpha,\beta}}{-i ([{\cal W}_0]_{\alpha,\alpha} + 
[{\cal W}_0]_{\beta,\beta}) +\frac{1}{2}  [{\cal G}   + {\cal R} ]_{\alpha,\alpha} + 
 [{\cal G}   + {\cal R} ]_{\beta,\beta}} \:\: .
 \label{ap.1.9}
 \end{eqnarray}
 \noindent
 \twocolumngrid
On iteratively solving Eqs.(\ref{ap.1.9}) it is possible to  compute $G_{{\rm int},n}$, as we discuss in the following.

\subsection{Charge current at the NS interface in the non-equilibrium steady state}
\label{i_ness}

We now employ the LME  approach  to compute $G_{{\rm int},n}$.  
To do so, we consider  the corresponding current operator, given by

\beq
J_{{\rm int},n} = - i \gamma \{\alpha_{{\cal N}}^\dagger \chi_n - \chi_n^\dagger \alpha_{{\cal N}} \} 
\:\: . 
\label{ap.2.1}
\eneq
\noindent
Using (the inverse of) Eqs.(\ref{lme.3},\ref{lme.5}) of the main text, we may readily express the average interface current 
at time $t$, $I_{{\rm int},n} (t)$, in terms of the matrix elements of $\theta (t)$ and of $\theta^A (t)$  as 

\bea
& & I_{{\rm int},n}  (t) = 2  \gamma \Im m {\rm Tr} [\rho (t) \alpha_{{\cal N}}^\dagger \chi_n ] \nonumber \\ 
& & = 2 \gamma \sum_{k,q}
\sqrt{\frac{2}{{\cal N} + 1}} \sin (k {\cal N} ) \nonumber \\
& & \times\Im m \{u_{q,n} \theta_{(L,k);(2K,q)} (t)  +
 [v_{q,n}]^*  \theta^A_{(L,k);(2K,q)} (t) \} 
\:\: . \nonumber \\
& &
\label{ap.2.2}
\eea
\noindent
While all the steps we went through so far rely on no specific approximations, in the following we make 
the assumption that the 2LRK is weakly coupled to the leads, so that we may 
stop the calculation of the right-hand side of Eq.(\ref{ap.2.2}) to second order in $\gamma$. Moreover, 
since we are interested in the current characterizing the non-equilibrium steady state, we use 
Eqs.(\ref{ap.1.9}) to compute the corresponding $\Theta$-matrix. In order to compute
$I_{{\rm int},n}$ to second order in $\gamma$, we approximate $\Theta_{\rm NESS}$ as
$\Theta_{\rm NESS} \approx \Theta_{{\rm NESS},0} + \gamma \Theta_{{\rm NESS},1}$. 
From Eq.(\ref{ap.2.2}) we therefore obtain, for the current through site $n$ in the NESS
to order $\gamma^2$

\bea
& & I_{{\rm int},n} = 2 \gamma^2   \sum_{k,q}
\sqrt{\frac{2}{{\cal N} + 1}} \sin (k {\cal N} ) \nonumber \\
& & \times \Im m \{u_{q,n} [\Theta_{{\rm NESS},1}]_{(L,k,1);(2K,q,1)}  \nonumber \\ 
& & +[v_{q,n}]^*   [\Theta_{{\rm NESS},1}]_{(L,k,-1);(2K,q,1)} \} 
\:\: .
\label{ap.2.2b}
\eea
\noindent
Integrating over the lead modes, we rewrite the right-hand side of Eq.(\ref{ap.2.2b}) as a sum
over contributions from the modes of the superconducting chain, that is

\beq
I_{{\rm int},n} = \sum_q I_{{\rm int},n}^q 
\;\; , 
\label{ap.2.2c}
\eneq
\noindent
with 

\begin{eqnarray}
 I_{{\rm int},n}^q  &=& 2 \gamma^2 \Gamma \:\frac{2}{{\cal N} +1} \:  \sum_k \Biggl\{ \frac{\sin^2 (k {\cal N} ) |u_{q,n}|^2
 (\bar{n}_{L,k} - \bar{n}_{2K,q})}{(\epsilon_{L,k}-\epsilon_{2K,q})^2 + \Gamma^2} \nonumber \\
 &+&  \frac{\sin^2 (k {\cal N} ) |b_{q,n}|^2
 (\bar{n}_{L,k} + \bar{n}_{2K,q}-1)}{(\epsilon_{L,k}-\epsilon_{2K,q})^2 + \Gamma^2} \Biggr\}
 \:\: , 
 \label{ap.2.2d}
 \end{eqnarray}
 \noindent
 with $\bar{n}_{L,k}$ and $\bar{n}_{2K,q}$ being respectively the average particle occupancy of 
 the lead level with momentum $k$ and of the superconductor level with momentum $q$. As the
 result in Eqs.(\ref{ap.2.2b},\ref{ap.2.2c}) is independent of the specific Hamiltonian for the 
 Kitaev chain, it keeps valid at any given realization of the disorder in the superconductor. 
 In addition, from Eq.(\ref{ap.2.2d})   we   find that the leading contribution in $\gamma$ to 
$I_{{\rm int},n}$ is given by the sum of two different terms, the former one determined
by the subgap, real fermionic modes (if any, that is, if the Kitaev chain is in a topological phase), 
the latter one, instead, being determined by the ``bulk'' states, at energy $\geq \Delta_{2K}$. 
In order to have a finite $I_{{\rm int},n}$, a voltage bias $V$ has to be applied between the 
metallic lead and the superconductor. Once we know $I_{{\rm int},n}$ as a function of 
$V$, we define the corresponding dc conductance at site $n$,  $G_{{\rm int},n}$, as 
$G_{{\rm int},n} = \lim_{V \to 0} \frac{\partial I_{{\rm int},n}}{\partial V}$. 

Just as $I_{{\rm int},n}$, also $G_{{\rm int},n}$ is typically written as a contribution from 
the localized edge states plus a contribution from the bulk states. 
We now prove that, as soon as $\Delta_{2K}$ is finite, regardless of whether the system is 
disordered, or not, in the zero-temperature limit
 the latter contribution is always $=0$ in the small-$\gamma$ limit. To evidence 
this point, we note that, following the same steps as in Ref.\cite{Guimaraes2016}, we get that the 
contribution to the total conductance arising from the delocalized (above the gap) states is 
given by

\beq
[G_{{\rm int},n}]_{\rm bulk} = \frac{4 \gamma^2}{\pi t^2} \: \sum_q \: \left[ \frac{|u_{q,n}|^2 |v_{q,n}|^2}{|u_{q,n}|^2 + 
|v_{q,n}|^2} \right]   \left[ \frac{ 2 t \Gamma }{\Gamma^2 + \epsilon_q^2}   \right]   
\;\; .
\label{ap.2.3}
\eneq
\noindent
In Eq.(\ref{ap.2.3}) we sum over the full Brillouin zone of the (isolated) Kitaev chain. For a long enough chain, we 
may trade the sum over $q$ for an integral over the energies, by introducing the density of state function
for the Kitaev chain, $\rho_{2K} (\epsilon)$, so that 
$\sum_q = \int \: d \epsilon \: \rho_{2K} (\epsilon)$. Let us assume that a generic realization of the disorder 
potential yields a finite gap $\Delta_{2K}$ in the spectrum of the superconducting chain. As long as 
$\Gamma \ll \Delta_{2K}$, we may approximate $\frac{\Gamma}{\Gamma^2 + \epsilon^2} \approx 
 \pi \delta (| \epsilon | - \Gamma )$. Thus, as there are no states  within the interval $-\Delta_{2K} \leq \epsilon \leq \Delta_{2K}$, 
 we readily find that $[G_{{\rm int},n}]_{\rm bulk} = 0$, $\forall n$. Let us, now, consider the contribution to the conductance
 arising from the subgap boundary states. 

\begin{figure}
\includegraphics[width=1.0\linewidth]{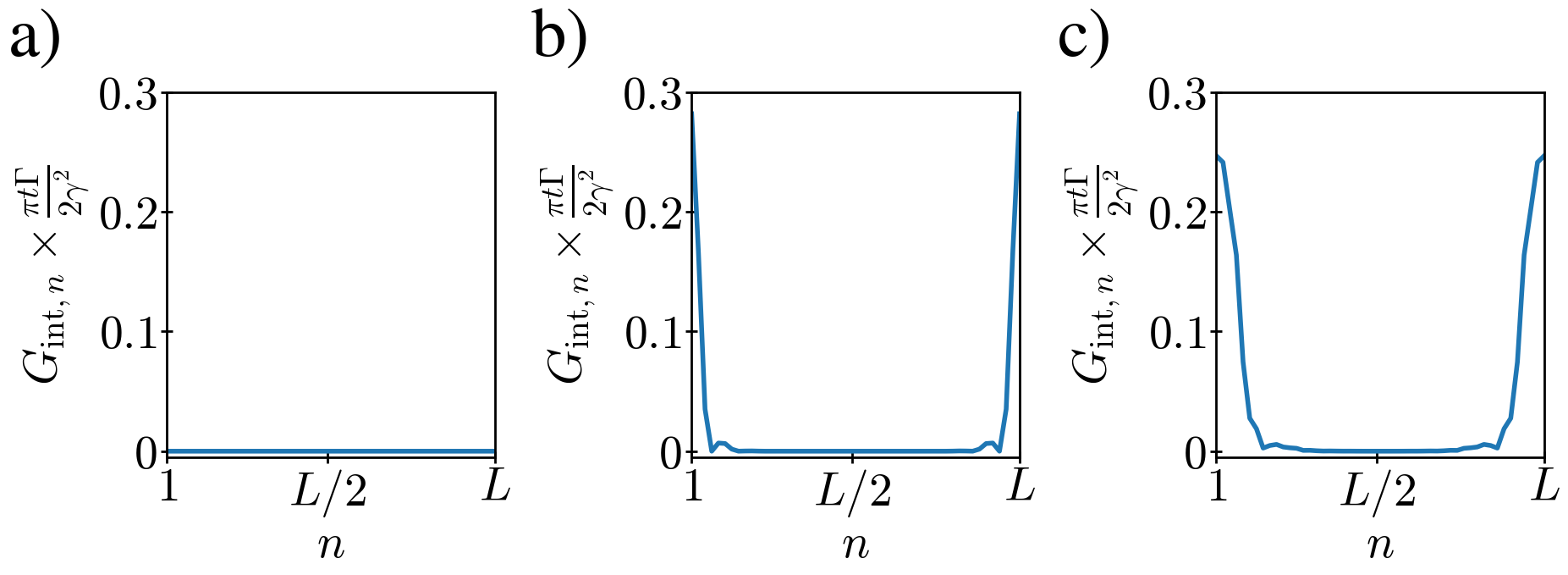}
\caption{$G_{{\rm int},n}$ computed in the clean 2LRK, for $L=50$ and with $g=1$ and, respectively, $\lambda_1=\lambda_2=0.5$ (panel {\bf a)}, 
corresponding to the topologically trivial phase, with no localized states), $\lambda_1=1.5,\lambda_2=0.3$ (panel {\bf b)}, corresponding 
to the topological phase with $\omega=1$), $\lambda_1=2.0,\lambda_2=1.5$
(panel {\bf c)} corresponding to the topological phase with $\omega=2$). 
 }
\label{condsite}
\end{figure}
 
 As a legitimate eigenstate of the Hamiltonian, each subgap state is described by a Bogoliubov-de Gennes wavefunction 
 in the form $(u_n,v_n)$, with $|u_{n}| = |v_{n}|$. In the large-$L$ limit (long Kitaev chain) we may neglect finite-size corrections to the localized
 state energies which, accordingly, are stuck at $\epsilon = 0$. In this limit and for $T\to 0$, the total contribution to the 
 dc conductance emerging from the localized boundary states is given by

\beq
[G_{{\rm int},n}]_{\rm loc} =\sum_{\alpha = {\rm loc} }  \:  \frac{4 \gamma^2 |u_{\alpha,n}|^2}{\pi t \Gamma} \: \left[ 
1 + \left(\frac{\Gamma}{2t} \right)^2 \right]   
\;\; ,
\label{ap.2.4}
\eneq
\noindent 
with the sum taken over the set of isolated states. 

From our discussion above, we find that, in the dc limit and at low enough temperature, only the subgap modes
(if any) contribute $G_{{\rm int},n}$. To evidence this result, in Fig.\ref{condsite} we plot the site-depending
conductance computed in the clean Kitaev chain, for $L=50$ and with $g=1$ and, respectively, $\lambda_1=\lambda_2=0.5$ (panel {\bf a)}, 
corresponding to the topologically trivial phase, with no localized states), $\lambda_1=1.5,\lambda_2=0.3$ (panel {\bf b)}, corresponding 
to the topological phase with a single subgap real fermionic mode at each endpoint of the chain), $\lambda_1=2.0,\lambda_2=1.5$
(panel {\bf c)} corresponding to two subgap real fermionic modes localized at each endpoint of the chain). From the 
three plots, we readily note that $G_{{\rm int},n}$ is practically 0 in the trivial phase and in the bulk of the topological phases. 
At variance, close to the boundaries of the chain, it jumps to a finite value ($\propto \frac{\gamma^2}{\pi t \Gamma} )$,
following the profile of $|u_{\alpha,n}|^2$ for the localized boundary states (note that for every subgap localized
state we have $|u_{\alpha,n}|^2 = |v_{\alpha,n}|^2$). Summing over $n$, the normalization condition for the localized state wavefunctions yields 

\beq
G_{\rm int} = 
\sum_{n=1}^L \: G_{{\rm int},n} = \frac{2 \gamma^2}{\pi t \Gamma} \left[1 + \left(\frac{\Gamma}{2t} \right)^2 \right] \omega 
\:\: ,
\label{ap.2.5}
\eneq
\noindent
with $\omega$ counting the number of subgap, real fermionic modes localized at each boundary of the chain. 
Thus, we conclude that a zero/finite value of $G_{\rm int}$ evidences the absence/presence of the subgap modes 
characterizing the topological phases of the Kitaev chain, marking the phase transition lines with abrupt (and quantized)
jumps in its value. As we discuss in the main text, on such an observation we ground our method to map out
the phase diagram of the disordered 2LRK by looking at the (disorder averaged) dc conductance as 
a function of the system parameters. 

\bibliography{biblio.bib}
\end{document}